\documentclass[conference]{IEEEtran}
\IEEEoverridecommandlockouts
\usepackage{cite}
\usepackage{amsmath,amssymb,amsfonts, mathtools}
\usepackage{algorithmic}
\usepackage{graphicx}
\usepackage{textcomp}
\usepackage{xcolor}
\def\BibTeX{{\rm B\kern-.05em{\sc i\kern-.025em b}\kern-.08em
    T\kern-.1667em\lower.7ex\hbox{E}\kern-.125emX}}
\usepackage{amsthm}
\newtheorem{theorem}[]{Theorem}

\newtheorem{proposition}[]{Proposition}

\newtheorem{definition}[]{Definition}

\usepackage{tikz,pgfplots}
\pgfplotsset{compat=1.18}
\usetikzlibrary{arrows.meta,calc,patterns}
\usepackage{multicol, multirow}
\usepackage{float}
\usepackage{booktabs}
\usepackage{subcaption}
\usepackage[font=small]{caption}
\usepackage{url}
\usepackage{enumerate}
\usepackage{hyperref}
\makeatletter
\newcommand{\linebreakand}{%
  \end{@IEEEauthorhalign}
  \hfill\mbox{}\par
  \mbox{}\hfill\begin{@IEEEauthorhalign}
}
\makeatother
\begin{document}

\title{Strategic Data Center Load Shifting: Implications for Market Efficiency and Transmission Value}

\author{Aron Brenner$^*$, Line Roald$^\dagger$, and Saurabh Amin$^*$\\
$^*$Laboratory for Information \& Decision Systems, MIT, Cambridge, MA, U.S.A.\\
$^\dagger$Department of Electrical \& Computer Engineering, UW-Madison, Madison, WI, U.S.A.

\thanks{Proofs of all theoretical results can be found in the supplementary material available at \url{https://doi.org/10.5281/zenodo.17636653}.}}

\maketitle

\begin{abstract}
Data center electricity use may reach 12\% of U.S. demand by 2030, alongside growing ability to shift workloads geographically in response to prices or carbon signals. We examine the system-level implications of such strategic flexibility using a bilevel two-zone model that couples economic dispatch with consumer cost minimization. Two market failures emerge. First, discontinuous price changes at generator capacity limits can induce flexible consumers to shift load in socially inefficient directions---for example, toward a higher-cost region to trigger a price drop elsewhere. Second, by positioning near capacity boundaries, consumers can counteract the marginal benefit of transmission expansion: although shadow prices suggest additional capacity is valuable, strategic consumers reoptimize to offset resulting flow changes, leaving dispatch and costs unchanged. We derive conditions under which these effects arise and show that conventional price signals can misrepresent system value in the presence of large spatially flexible loads.
\end{abstract}


\section{Introduction}

Data center growth is rapidly reshaping electricity demand. In the U.S., data center consumption may more than triple from 2023 to 2030, reaching $\sim$12\% of total demand~\cite{mckinsey2024}. Because large operators can migrate workloads, these facilities act not only as major loads but also as sources of spatial and temporal flexibility---capabilities already used to reduce costs or emissions of hyperscale computing~\cite{google_env_report,Meta_sust_report,ZhengEtal2020_joule,radovanovic2022carbon}.

A common assumption is that such flexibility should improve system efficiency: by shifting load toward locations with lower locational marginal prices (LMPs), flexible consumers provide benefits similar to transmission---reducing congestion and better utilizing low-cost generation~\cite{zhang2020flexibility}. This view, however, relies on consumers behaving as price-takers. In contrast, empirical studies of electricity markets document that some large participants strategically exert market power, often at a net cost to social welfare~\cite{Borenstein2000, hortacsu2005understanding}. As data center loads grow large enough to influence prices, incentives emerge for strategic load shifting that undermines system efficiency.

These observations motivate two central questions:
\begin{quote}
\begin{enumerate}
\item \textit{When can strategic load shifting \textbf{reduce} system efficiency?}
\item \textit{How do such shifting incentives reshape the effective \textbf{value of transmission}?}
\end{enumerate}
\end{quote}
The second question follows from the first: if flexible load can change dispatch outcomes, the benefits of transmission expansion may depend on consumers' strategic responses.

Prior work exploring the benefits of flexibility often assumes centralized coordination~\cite{zhang2020flexibility,knittel2025} or treats consumers as \textit{price- or emissions-takers}~\cite{radovanovic2022carbon,SouzaEtal2023}, and therefore does not capture strategic interactions by large flexible loads. While recent work has begun to identify counterintuitive outcomes from decentralized load shifting with respect to emissions~\cite{GorkaEtal2024}, the mechanisms by which flexible consumers influence prices and congestion---and the resulting implications for transmission value---remain underexplored.

We address this gap using a bilevel model in which a flexible consumer chooses spatial load allocation while the system operator clears the resulting market. In a two-zone system, we derive closed-form conditions under which a flexible consumer's incentives align with or diverge from system objectives. Our contributions are:

\textbf{1. Strategic misalignment.} We show that decentralized shifting can raise system costs even as flexible consumers lower their own. Because consumers pay marginal-cost-based prices for load while the system incurs marginal costs only for incremental dispatch, consumers can benefit from positioning near generator capacity limits to depress prices (Theorem~\ref{thm:externalities}).

\textbf{2. Zero bilevel marginal value of transmission.} At these strategic equilibria, incremental transmission expansion has zero marginal value: flexible consumers shift load one-for-one with new line capacity, offsetting its effect and leaving dispatch unchanged, even as shadow prices remain positive.

\textbf{3. Implications for planning and design.} Because marginal pricing can induce strategic behavior that departs from socially optimal outcomes and biases transmission investment signals, our findings point to a potential need to refine planning and market models to better account for strategic flexible demand.

\section{Two-Zone Bilevel Model}
We model the interaction between a spatially flexible electricity consumer and a system operator as a bilevel optimization problem. The consumer (upper level) chooses how to allocate load across two zones to minimize procurement costs, while the system operator (lower level) clears the market given a baseline system load and the consumer's allocation decision.

\textbf{Generation and network structure.} Consider a power system with two zones, $A$ and $B$, each containing a merit-order supply stack of generation resources. Each generator $g \in \mathcal{G}_A \cup \mathcal{G}_B$ has capacity $0 \leq p_g \leq P_g$ (MW) and constant marginal cost $C_g \geq 0$ (\$/MWh). Let $F \geq 0$ denote the transmission capacity between the zones, and let $f \in [-F, F]$ denote power flow from $A$ to $B$.

Generators in each zone are ordered by strictly increasing marginal cost. For a given dispatch, let $g_i$ denote the marginal generator at zone $i \in \{A,B\}$, with $g_i \pm 1$ denoting the adjacent generators in the merit order when they exist. This strict ordering is without loss of generality, as generators with identical marginal costs can be aggregated into a single unit.

\textbf{Load allocation.} Each zone has two types of loads: (i) \emph{Inflexible base load}: $b_A,\, b_B \geq 0$ (MWh), which cannot be shifted;  (ii) \emph{Flexible consumer load}: Initially allocated as $x_A,\, x_B \geq 0$ (MWh). Importantly, the flexible consumer can shift load between zones. Let $\delta \in [\underline{\delta}, \overline{\delta}]$ denote the \emph{net load shift from zone $A$ to zone $B$}, where $\delta > 0$ (resp. $\delta <0$) means net shift from $A$ to $B$ (resp. from $B$ to $A$). The bounds $\underline{\delta}$ and $\overline{\delta}$ capture operational constraints imposed by the flexible consumer (e.g., service constraints). After shifting, the realized flexible loads are $x_A - \delta$ at zone $A$ and $x_B + \delta$ at zone $B$.

\subsection{Lower Level: Economic Dispatch} Given a load shift $\delta$, the system operator solves an economic dispatch to minimize \textit{total generation cost}:
\begin{definition}[Market Clearing Problem]
    \label{def:market_clearing}
    \begin{subequations}
    \label{eq:dispatch}
    \begin{align}
        G(\delta, F) := \min_{p,\,f} \quad & \sum_{g \in \mathcal{G}_A \cup \mathcal{G}_B} C_g\,p_g \label{eq:dispatch_obj}\\
        \mathrm{s.t.} \quad & \sum_{g \in \mathcal{G}_A} p_g = b_A + x_A - \delta + f \quad (\lambda_A) \label{eq:balance_A}\\
        & \sum_{g \in \mathcal{G}_B} p_g = b_B + x_B + \delta - f \quad (\lambda_B) \label{eq:balance_B}\\
        & -F \leq f \leq F \hspace{1.8cm} (\mu^-, \mu^+) \label{eq:transmission}\\
        & 0 \leq p_g \leq P_g, \hspace{1.1cm} \forall g \in \mathcal{G}_A \cup \mathcal{G}_B \label{eq:generation}
    \end{align}
    \end{subequations}
\end{definition}
The dual variables have standard economic interpretations: $\lambda_A, \lambda_B$ denote the LMPs at zones $A$ and $B$; $\mu^+, \mu^-$ the shadow price of upper/lower transmission limits. Importantly, LMPs equal the marginal cost of the marginal generator serving each zone.

We assume that all demand is inelastic and must be fully served (i.e., no load shedding) with willingness-to-pay exceeding the highest marginal generation cost. Under this standard assumption, minimizing the system objective $G(\delta,F)$ is equivalent to \textit{maximizing social welfare} \cite{schweppe1988}.

\subsection{Upper Level: Flexible Consumer Problem}
The flexible consumer chooses a load shift $\delta$ to minimize their own \textit{procurement cost}; we assume the consumer has perfect information about parameters of the market clearing problem \eqref{eq:dispatch}.
\begin{definition}[Consumer Cost Minimization]
    \label{def:consumer_problem}
    Let $\lambda_A(\delta, F)$ and $\lambda_B(\delta, F)$ denote the LMPs resulting from dispatch problem \eqref{eq:dispatch}. The consumer's cost is:
    \begin{equation}
        P(\delta, F) := \lambda_A(\delta, F) (x_A - \delta) + \lambda_B(\delta, F) (x_B + \delta)
    \label{eq:consumer_cost}
    \end{equation}
    The consumer solves:
    \begin{equation}\label{eq:consumer_opt}
        \min_{\underline{\delta}\leq\delta\leq\overline{\delta}} \quad P(\delta, F)
    \end{equation}
\end{definition}
Unlike the system objective $G(\delta, F)$, which is convex and continuous in $\delta$, the consumer objective $P(\delta, F)$ exhibits two key properties:
\begin{enumerate}
    \item \textbf{Nonconvexity}: $P(\delta, F)$ is generally nonconvex due to the dependence on endogenous prices $\lambda_i(\delta, F)$    
    \item \textbf{Discontinuities}: When $\delta$ crosses a threshold where the marginal generator changes, the LMPs jump discontinuously. This causes $P(\delta, F)$ to have upward (resp. downward) jumps at these thresholds where more (resp. less) expensive generators become marginal.
\end{enumerate}
These discontinuities are the fundamental source of misalignment between private and social incentives, which we analyze in Sec.~\ref{sec:results}.

\subsection{Objective Alignment}

\begin{definition}[Objective Alignment]
    \label{def:alignment}
    Fix a reference point $\delta_0$ (e.g., $\delta_0=0$ for no shifting). Define the incremental changes:
    \begin{align}
    \Delta P(\delta) &:= P(\delta, F) - P(\delta_0, F)\\
    \Delta G(\delta) &:= G(\delta, F) - G(\delta_0, F)
    \end{align}
    We say the system and consumer objectives are \textbf{misaligned} with respect to shift $\delta$ if: 
    \begin{equation}
        \Delta P(\delta) < 0 \quad \quad \text{and} \quad \quad \Delta G(\delta) > 0,
        \label{eq:alignment_condition}
        \end{equation}
    i.e., if the flexible consumer prefers a shift that increases system cost, and \textbf{aligned} otherwise. 
\end{definition}

While deliberately stylized, our two-zone framework captures key mechanisms of price formation and congestion while remaining analytically tractable. Setting $F=0$ models disconnected markets, isolating how differences in local supply stacks can create misalignment between system and consumer objectives, while $F>0$ represents interconnected regions, highlighting how spatially flexible demand interacts with congestion and marginal pricing to distort transmission value and investment signals.

\section{Disconnected Electricity Markets}\label{sec:results}
In the following section, we study the case of two disconnected electricity markets (i.e., assuming $F=0$) and identify conditions under which system generation cost and flexible consumer cost objectives are aligned or misaligned. For now, we adopt the notation $G(\delta) \coloneqq G(\delta,0)$ and $P(\delta) \coloneqq P(\delta,0)$.

\subsection{Basis Changes and Discontinuities}
When load shifts are large enough to change marginal generators, the flexible consumer cost function develops discontinuities that can lead to misalignment. Proposition~\ref{thm:objective_relation} characterizes how system and flexible consumer objectives change at basis changes of \eqref{eq:dispatch}, denoted by $\tau_1$ and $\tau_2$.

\begin{proposition}\label{thm:objective_relation}
Without loss of generality, consider a positive load shift $\delta > 0$ from $A$ to $B$. Let $p_{g_A}$ and $p_{g_B}$ denote the pre-shift dispatch levels of the marginal generators $g_A$ and $g_B$, respectively. We assume $p_{g_A} \neq P_{g_B} - p_{g_B}$ so that exactly one marginal generator changes first as $\delta$ increases. Two cases then arise, depending on which zone's marginal generator changes first.

\smallskip
\noindent\textbf{Case 1.} (Marginal generator at $B$ changes first).
Suppose
\begin{align*}
    P_{g_B} - p_{g_B} < p_{g_A}.
\end{align*}
Let $\tau_1 := P_{g_B} - p_{g_B}$ be the threshold where generator $g_B$ reaches capacity. Then, for $\delta \leq \tau_1$:
\begin{align*}
    \Delta G(\delta) =
    \begin{cases}
        \Delta P(\delta), & \delta \in [0, \tau_1), \\[2pt]
        \Delta P(\delta) - \kappa_1, & \delta = \tau_1,
    \end{cases}
\end{align*}
where the discontinuity $\kappa_1$ is given by
\begin{align*}
    \kappa_1 \coloneqq (C_{g_{B+1}} - C_{g_B})(x_B + \tau_1) > 0.
\end{align*}

\noindent\textbf{Case 2.} (Marginal generator at $A$ changes first).
Suppose
\begin{align*}
    p_{g_A} < P_{g_B} - p_{g_B}.
\end{align*}
Let $\tau_2 := p_{g_A}$ be the threshold where generator $g_A$ is fully de-committed. Then, for $\delta \leq \tau_2$:
\begin{align*}
    \Delta G(\delta) =
    \begin{cases}
        \Delta P(\delta), & \delta \in [0, \tau_2), \\[2pt]
        \Delta P(\delta) + \kappa_2, & \delta = \tau_2,
    \end{cases}
\end{align*}
where the discontinuity $\kappa_2$ is given by
\begin{align*}
    \kappa_2 \coloneqq (C_{g_A} - C_{g_{A-1}})(x_A - \tau_2) > 0.
\end{align*}
\end{proposition}

Interpreting Proposition~\ref{thm:objective_relation}, we first note that, for $\delta$ smaller than the first basis-change threshold (i.e., before either $\tau_1$ or $\tau_2$ is reached):
\begin{align*}
    \Delta G(\delta) = \Delta P(\delta) = (C_{g_B} - C_{g_A})\,\delta.
\end{align*}
In other words, when load shifts are small enough that the optimal basis of the dispatch problem does not change, system and consumer objectives are perfectly aligned: both benefit linearly from shifting toward the lower-cost region.

Once a basis change occurs, however, the consumer pays the new marginal price on all load at affected locations while the system only pays incrementally. This yields discontinuities in the consumer objective at $\delta = \tau_1$ or $\delta = \tau_2$, creating incentives for flexible consumers to depress prices by strategically shifting load to alter which generators are marginal in zones $A$ and $B$, even when such shifts increase total system cost.

\subsection{Strategic Positioning and Misalignment}
Theorem~\ref{thm:externalities} formalizes the conditions under which these discontinuities induce load shifting outcomes that raise system costs.

\begin{theorem}[Externalities at basis changes]\label{thm:externalities}
Without loss of generality, consider a positive load shift $\delta=\tau>0$ from $A$ to $B$ that is \emph{exactly} the minimal shift required to change the optimal basis of \eqref{def:market_clearing} (i.e., to induce a change in the marginal generator at either $A$ or $B$). Let $p_{g_A}$ and $p_{g_B}$ denote the pre-shift dispatch levels of the marginal generators at $A$ and $B$. 

The system and consumer objectives are \emph{misaligned} with respect to $\delta$ if and only if all of the following conditions hold:

\begin{enumerate}
    \item[(i)] The minimal basis-changing shift causes \emph{de-commitment} of the marginal generator at $A$:
    \begin{align*}
        \tau = p_{g_A} < P_{g_B} - p_{g_B}.
    \end{align*}

    \item[(ii)] Before the shift, the marginal generator at $B$ is strictly more expensive:
    \begin{align*}
        C_{g_B} > C_{g_A}.
    \end{align*}

    \item[(iii)] The downward jump in the price at $A$ (from switching $g_A$ to $g_{A-1}$) offsets the increase in price exposure at $B$:
    \begin{align*}
        (C_{g_A} - C_{g_{A-1}})(x_A - \tau) > (C_{g_B} - C_{g_A})\,\tau.
    \end{align*}
\end{enumerate}
\end{theorem}

In other words, a flexible consumer is incentivized to shift load against LMP differences, and consequently raise generation costs, when the immediate cost drop from de-committing a marginal generator dominates the incremental cost of shifting load toward a higher-priced region.

\subsection{Illustrative Numerical Example}\label{sec:theorem_1_example}
\begin{figure}[htbp]
    \centering
    \includegraphics[width=\linewidth]{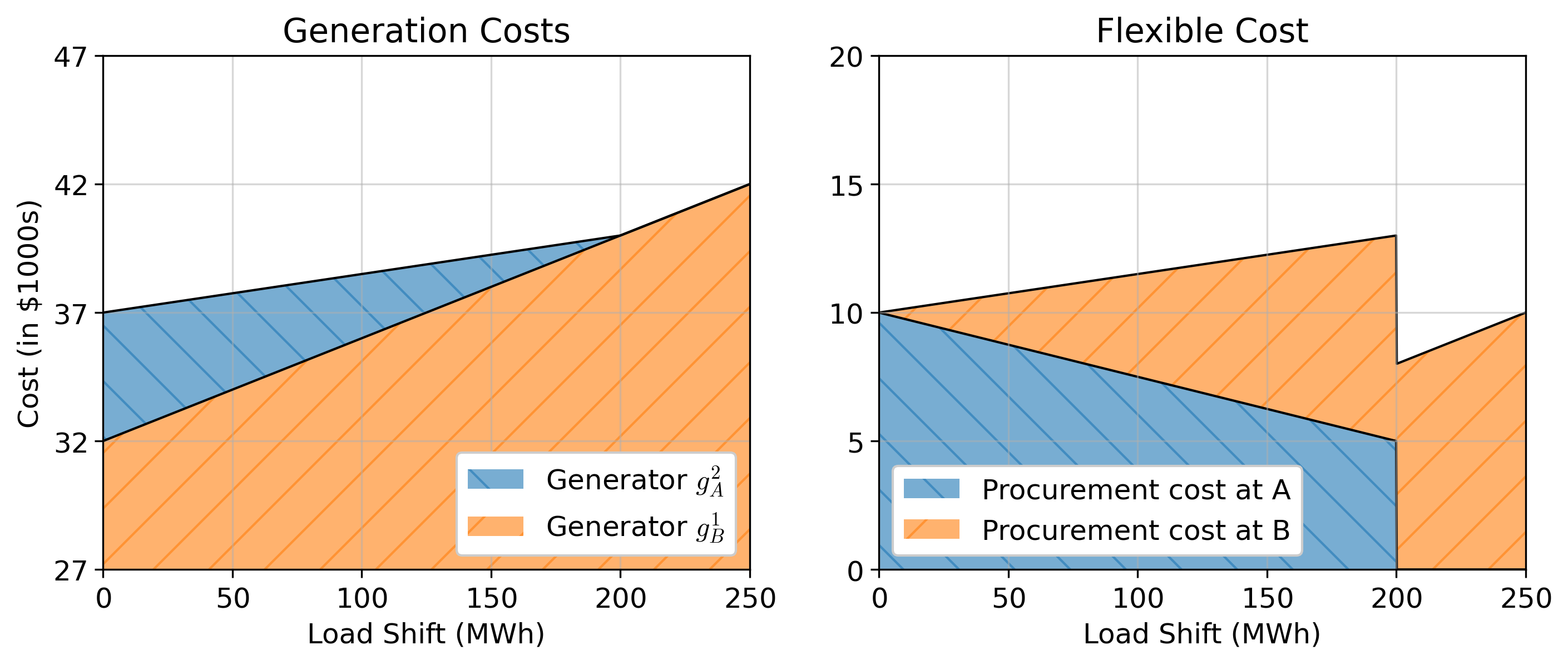}
    \caption{Comparison of system generation costs (left) and flexible consumer costs (right) for a range of shifts $\delta \in [0, 250]$.}
    \label{fig:theorem_1}
\end{figure}

We illustrate this phenomenon in Fig.~\ref{fig:theorem_1} using a simple two-zone example in which flexible load accounts for 20\% of total demand. The system parameters are:

\textbf{Base loads:} $b_A=b_B=800$ MWh

\textbf{Flexible loads:} $x_A=400$ MWh and $x_B=0$ MWh

\textbf{Zone $\boldsymbol{A}$ generators:}
\begin{enumerate}
    \item $g_A^1$: 1,000 MW at \$0/MWh
    \item $g_A^2$: 200 MW at \$25/MWh
\end{enumerate}

\textbf{Zone $\boldsymbol{B}$ generators:}
\begin{enumerate}
    \item $g_B^1$: 1,200 MW at \$40/MWh
\end{enumerate}
In this configuration, a flexible consumer minimizes procurement cost by shifting half of their load from $A$ to $B$ so that generator $g_A^2$ is no longer marginal. This reduces the LMP at $A$ from \$25/MWh to \$0/MWh. However, the resulting load shift forces a larger share of energy to be supplied by $g_B^1$ at \$40/MWh, increasing total system generation costs. This occurs because the cost gap between $g_A^1$ and $g_A^2$ is large relative to the gap between $g_A^2$ and $g_B^1$, creating an incentive for strategic shifting even though it raises system costs.

\section{Transmission-Coupled Markets}
We now introduce transmission coupling ($F > 0$), which creates additional strategic interactions beyond those in Section~\ref{sec:results}. When load is inflexible, the short-run marginal value of transmission---defined by $-\frac{d}{dF}G$, or the reduction in cost per unit of added capacity---is conventionally understood to be the shadow price of the binding transmission constraint \cite{hogan1992contract}. With flexible load, however, locational prices influence the spatial allocation of demand, meaning that transmission and load shifting are endogenously coupled. This raises the question of how strategic load shifting affects the marginal value of increasing transmission capacity.

\subsection{Marginal Value of Transmission Under Flexible Load}
Theorem~\ref{thm:marginal_value_transmission} addresses this question in a regime where flexible load optimally positions at a basis change by de-committing the otherwise-marginal generator at $A$ (see Theorem~\ref{thm:externalities}). Under these conditions, transmission upgrades no longer yield the cost reductions suggested by the shadow price.

\begin{theorem}[Zero bilevel marginal value of transmission]\label{thm:marginal_value_transmission}
Fix initial transmission capacity $F_0 \geq 0$ and let $\delta_0 := \delta^*(F_0)$ be the consumer-optimal load shift. Suppose the following conditions hold:

\begin{enumerate}[(i)]
    \item[(A1)] Marginal generation costs satisfy $C_{g_A} < C_{g_{A+1}} < C_{g_B}$ and the transmission line is congested from $A$ to $B$:
    \begin{align*}
        \lambda_A = C_{g_A} < C_{g_B} = \lambda_B, \qquad f = F_0.
    \end{align*}

    \item[(A2)] Strategic positioning is profitable:
    \begin{align*}
       (C_{g_{A+1}} - C_{g_A})\,(x_A - \delta_0) > (C_{g_B} - C_{g_{A+1}})\,\delta_0,
    \end{align*}
    so that $g_A$ is being dispatched at level $p_{g_A} = P_{g_A}$.

    \item[(A3)] The marginal generator $g_B$ has downward headroom: $p_{g_B} > 0$.
\end{enumerate}
Define the threshold transmission capacity:
\begin{align*}
    \Gamma := \left(\frac{C_{g_{A+1}} - C_{g_A}}{C_{g_B} - C_{g_A}}\right) x_A - \delta_0
\end{align*}
and the upper limit:
\begin{align*}
F_1 := F_0 + \min\left\{\Gamma,\,\overline{\delta} - \delta_0\right\}
\end{align*}
Then for all $F \in [F_0, F_1)$:

\begin{enumerate}[(i)]
    \item The transmission constraint remains binding with positive shadow price:
    \begin{align*}
    \mu^+ = C_{g_B} - C_{g_A} > 0.
    \end{align*}

    \item The marginal value of transmission w.r.t. the system objective is zero under load shifting:
    \begin{align*}
    -\frac{d}{dF}\,G(\delta^*(F),F) = 0.
    \end{align*}

    \item The marginal value of transmission w.r.t. the flexible consumer objective is negative under load shifting:
    \begin{align*}
    -\frac{d}{dF}\,P(\delta^*(F),F) = C_{g_A} - C_{g_B} < 0.
    \end{align*}
\end{enumerate}
\end{theorem}

In this regime, the shadow price $\mu^+$ no longer reflects the marginal value of transmission capacity. If all load were held fixed, increasing $F$ would allow zone $B$ to replace higher-cost local generation with lower-cost supply from zone $A$, yielding a reduction in system cost. However, at the equilibrium described in Theorem~\ref{thm:marginal_value_transmission}, the flexible load best response adjusts to keep generator $g_A$ marginal in zone $A$. As $F$ increases, the optimal load shift $\delta^*(F)$ increases one-for-one, leaving the effective net transfer between zones unchanged.

As a consequence, the dispatch and total system cost remain constant even as transmission capacity expands: the \emph{system-level marginal value of transmission is zero}. Meanwhile, flexible consumer costs increase, since maintaining $g_A$ as the marginal generator in zone $A$ requires allocating more load to the higher-cost zone $B$. Thus, although transmission expansion appears beneficial when evaluated through its shadow price, this apparent benefit is fully offset by strategic load shifting.

\subsection{Illustrative Numerical Example}
To illustrate the interaction between load shifting and transmission capacity, we consider a system with initial base and flexible loads identical to those in Sec.~\ref{sec:theorem_1_example} and with the following generation resources.

\textbf{Zone $\boldsymbol{A}$ generation resources:}
\begin{enumerate}
    \item $g_1^A$: 400 MW capacity at \$0/MWh
    \item $g_2^A$: 600 MW capacity at \$25/MWh
    \item $g_3^A$: Unlimited capacity at \$60/MWh
\end{enumerate}

\textbf{Zone $\boldsymbol{B}$ generation resources:}
\begin{enumerate}
    \item $g_1^B$: 200 MW capacity at \$0/MWh
    \item $g_2^B$: 400 MW capacity at \$40/MWh
    \item $g_3^B$: Unlimited capacity at \$70/MWh
\end{enumerate}

\begin{figure}[t]
    \centering
    \includegraphics[width=\linewidth]{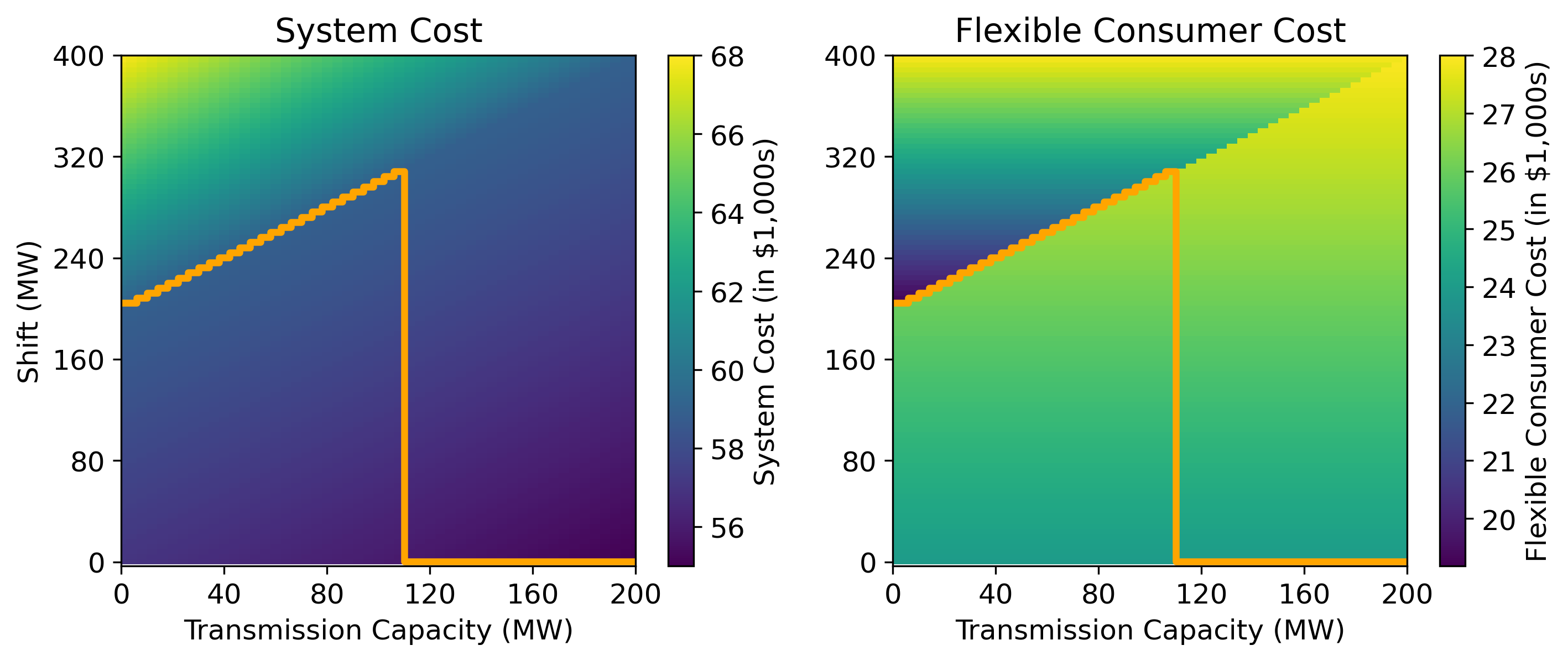}
    \caption{System costs (left) and flexible consumer costs (right) evaluated over a range of load shifts $\delta$ and transmission capacities $F$.}
    \label{fig:heatmaps}
\end{figure}

We solve the market-clearing problem \eqref{def:market_clearing} over $\delta\in[0,400]$ and $F\in[0,200]$, and plot the resulting objective values $G(\delta,F)$ and $P(\delta,F)$ as heatmaps (Fig.~\ref{fig:heatmaps}). System operating conditions and cost metrics appear in Table~\ref{tab:results}.

For all $F$, the system-optimal load shift is $\delta=0$, which allows full use of cheaper generation at $A$ and avoids dispatching the high-cost unit $g_B^3$. However, when $F < \Gamma$ (107 MW), flexible consumers reduce their own costs by shifting load to $B$ to avoid dispatching $g_A^3$, thereby raising the price at $A$ from \$25/MWh to \$60/MWh. This blocks the system operator from serving $B$ with cheaper power from $g_A^3$, increasing total system cost (see $F=0$ and $F=100$ in Table~\ref{tab:results}).

This illustrates the \emph{misalignment} in objectives described in Theorem~\ref{thm:marginal_value_transmission}: the cost gap between marginal units within zone $A$ (between $g_A^3$ and $g_A^2$) is large relative to cross-zone differences, giving flexible load an incentive to shift even though it raises system cost. For $F \ge \Gamma$, this incentive disappears because serving load in zone $B$ becomes more expensive than depressing prices in zone $A$, restoring alignment between system and consumer objectives (see $F=200$).

\begin{table}[b]
    \centering
    \renewcommand{\arraystretch}{1.1}
\begin{tabular}{lccc}
\toprule
& \multicolumn{3}{c}{Transmission Cap. (MW)} \\
\textbf{Metric} & $F=0$ & $F=100$ & $F=200$ \\
\midrule
\multicolumn{4}{l}{\textit{\textbf{No Load Shifting ($\delta = 0$)}}} \\
System Cost (\$) & 57{,}000 & 56{,}000 & 55{,}000 \\
Flexible Consumer Cost (\$) & 24{,}000 & 24{,}000 & 24{,}000 \\
LMPs $(\lambda_A,\lambda_B)$ (\$/MWh) & (60,\,70) & (60,\,70) & (60,\,70) \\
Marginal Units & $g_A^3,\, g_B^3$ & $g_A^3,\, g_B^3$ & $g_A^3,\, g_B^3$ \\
\midrule
\multicolumn{4}{l}{\textit{\textbf{Optimal Load Shifting}}} \\
Optimal $\delta^*$ (MW) & 200 & 300 & 0 \\
System Cost (\$) & 59{,}000 & 59{,}000 & 55{,}000 \\
Flexible Consumer Cost (\$) & 19{,}000 & 23{,}500 & 24{,}000 \\
LMPs $(\lambda_A,\lambda_B)$ (\$/MWh) & (25,\,70) & (25,\,70) & (60,\,70) \\
Marginal Units & $g_A^2,\, g_B^3$ & $g_A^2,\, g_B^3$ & $g_A^3,\, g_B^3$ \\
\midrule
\multicolumn{4}{l}{\textit{\textbf{Impact of Load Shifting}}} \\
$\Delta$ System Cost (\$) & +2{,}000 & +3{,}000 & 0 \\
$\Delta$ Flexible Cost (\$) & -5{,}000 & -500 & 0 \\
\bottomrule
\end{tabular}
    \caption{Dispatch decisions and cost metrics for $F=0,100,200$.}
    \label{tab:results}
\end{table}

\begin{figure}[tb]
    \centering
    \includegraphics[width=\linewidth]{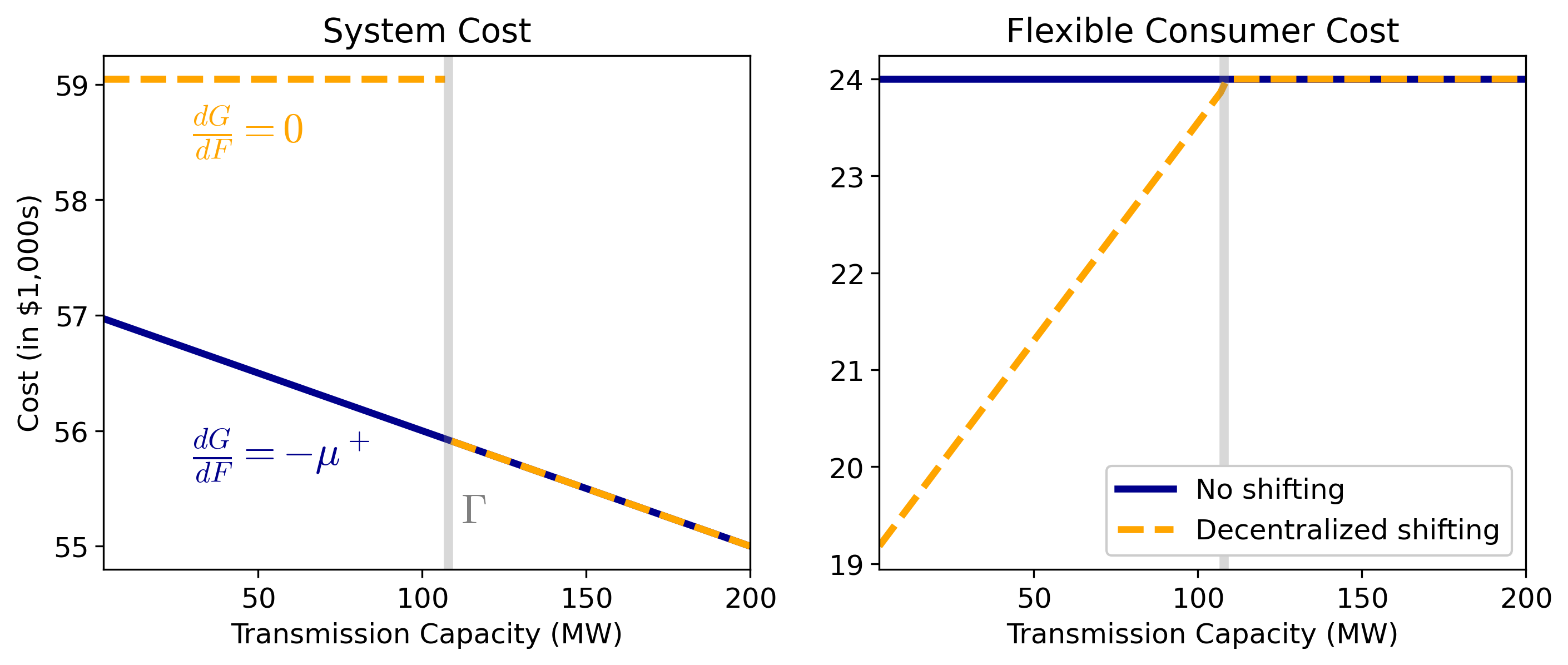}
    \caption{Comparison of system (left) and flexible consumer costs (right) without load shifting (blue) and with strategic load shifting (orange) for a range of transmission capacities.}
    \label{fig:costs}
\end{figure}

The implications of Theorem~\ref{thm:marginal_value_transmission} are further illustrated in Fig.~\ref{fig:costs}, which compares system and flexible consumer costs with respect to $F$ both with and without load shifting. For $F < \Gamma$, strategic load shifting exactly offsets the benefits of transmission expansion, yielding no change in system cost while raising costs for the flexible consumer. For $F \geq \Gamma$, this offset vanishes and transmission capacity again provides system value.

\section{Conclusion}
We develop an analytical framework showing how price-making behavior by flexible consumers can reduce market efficiency and distort standard investment signals. By deriving closed-form expressions for private and social objectives, we characterize when strategic load shifting aligns with, or diverges from, system-optimal dispatch, and show that conventional measures of transmission value---such as shadow prices---may fail when demand adjusts endogenously. These results underscore the need to account for strategic load response in capacity expansion models and market design.

While our analysis focuses on a two-zone system with perfectly informed consumers for analytical tractability, the underlying mechanism---price discontinuities induced by basis changes in dispatch---extends naturally to general networked systems. In larger networks, flexible loads may similarly position near generator or transmission capacity limits to induce discrete price changes. In practice, such strategic positioning depends on consumers' ability to anticipate price responses, and imperfect information or forecasting error may therefore limit how frequently such behavior arises. Characterizing the resulting misalignment in meshed networks and under imperfect information is an important direction for future work.

\section*{Acknowledgments}
The authors would like to thank Nathan Engelman Lado, Rahman Khorramfar, and Daniel Shen for their insights as well as the MIT Energy Initiative and MIT Climate Grand Challenges, which contributed to the development of this work. The authors also acknowledge the use of ChatGPT for assistance with editing this manuscript.

\bibliographystyle{IEEEtran}
\bibliography{bibliography}

\end{document}